\documentclass[aps,twocolumn,prb]{revtex4}
\usepackage{amssymb}
\usepackage{amsmath}
\usepackage{graphicx}
\usepackage{ulem}

\def\ttgq{\times 2e^2/h}
\def\tgq{2e^2/h}

\def\vsd{V_{\mathrm{sd}}}
\def\vdc{V_{\mathrm{dc}}}
\def\vac{V_{\mathrm{ac}}}
\def\vgone{V_{\mathrm{g1}}}
\def\vgtwo{V_{\mathrm{g2}}}

\def\Hethree{^3\mathrm{He}}
\def\Hefour{^4\mathrm{He}}
\def\pt7{$0.7\times2e^2/h$}

\def\tauns{\tau_{n,\sigma}}
\def\kb{k_{\mathrm{B}}}
\def\te{T_{e}}
\def\Sipart{S_{I}^{\mathrm{P}}}

\def\Tfridge{T_{\mathrm{fridge}}}
\def\gx{G_{X}}
\def\iq{I_{\mathrm{h}}}
\def\vdq{V_{\mathrm{h,d}}}
\def\vdsq{V_{\mathrm{h,ds}}}
\def\vdac{V_{\mathrm{dac}}}
\def\Reff{R_{\mathrm{eff}}}
\def\nVprH{\mathrm{nV}/\sqrt{\mathrm{Hz}}}
\def\tint{\tau_\mathrm{int}}
\def\NF{\mathcal{N}}

\begin{document}
\title{A system for measuring auto- and cross-correlation of
current noise at low temperatures}
\author{L.\ DiCarlo\footnote[1]{These authors contributed equally to this work.}, Y.\ Zhang\footnotemark[1], D.\ T.\ McClure\footnotemark[1], C.\ M.\ Marcus}
\affiliation{Department of Physics, Harvard University, Cambridge,
Massachusetts 02138, USA}
\author{L.~N.~Pfeiffer, K.~W.~West}
\affiliation{Bell Laboratories, Lucent Technologies, Murray Hill, NJ
07974, USA}
\date{\today}

\begin{abstract}
We describe the construction and operation of a two-channel noise
detection system for measuring power and cross spectral densities
 of current fluctuations near 2~MHz in electronic devices at low
temperatures. The system employs cryogenic amplification and fast-Fourier-transform based spectral measurement. The gain and
electron temperature are calibrated using Johnson noise
thermometry. Full shot noise of $100~\mathrm{pA}$ can be resolved
with an integration time of $10~\mathrm{s}$. We report a
demonstration measurement of bias-dependent current noise in a gate
defined GaAs/AlGaAs quantum point contact.
\end{abstract}
\maketitle

Over the last decade, measurement of electronic noise in mesoscopic conductors
has successfully probed quantum statistics, chaotic scattering and
many-body effects \cite{Blanter00,Martin05}. Suppression of shot noise
below the Poissonian limit has been observed in a wide range of
devices, including quantum point contacts~\cite{Reznikov95,Kumar96,Liu98},
diffusive wires~\cite{Steinbach96, Henny99a}, and quantum
dots~\cite{Oberholzer01}, with good agreement between experiment and
theory. Shot noise has been used to measure quasiparticle charge
in strongly correlated systems, including the fractional
quantum hall regime\cite{de-Picciotto97,Saminadayar97}
and normal-superconductor interfaces\cite{Jehl00}, and to investigate regimes
where Coulomb
interactions are strong, including coupled localized states in mesoscopic 
tunnel junctions~\cite{Safonov03} and quantum dots in the sequential tunneling~\cite{Gustavsson05} 
and cotunneling~\cite{Onac06} regimes. Two-particle interference not evident 
in dc transport has been
investigated using noise in an electronic beam splitter \cite{Liu98}. 

Recent theoretical
work\cite{Martin02,Samuelsson04,Beenakker04,Lebedev05}
proposes the detection of electron entanglement via violations of Bell-type
inequalities using  cross-correlations of current noise between different leads.
Most noise measurements have investigated either noise
auto-correlation\cite{Reznikov95,Steinbach96,de-Picciotto97,Schoelkopf97,Liu98,Spietz03,Onac06}
or cross-correlation of noise in a common
current\cite{Kumar96,Glattli97,Henny99a,Sampietro99,Oberholzer01,Safonov03},
with only a
few experiments\cite{Henny99,Oberholzer06,Oliver99} investigating
cross-correlation between two distinct currents.
Henny \textit{et al.}\cite{Henny99} and Oberholzer \textit{et al.}\cite{Oberholzer06}
measured noise cross-correlation in the acoustic frequency range (low kHz) 
using room temperature amplification and a commercial fast Fourier
transform (FFT)-based spectrum analyzer. Oliver \textit{et
al.}\cite{Oliver99} measured cross-correlation in the low MHz
using cryogenic amplifiers and analog power detection with hybrid
mixers and envelope detectors.

In this paper, we describe a two-channel
noise detection system for simultaneously measuring power spectral
densities and cross spectral density of current fluctuations in
electronic devices at low temperatures. Our approach combines
elements of the two methods described above: cryogenic amplification
at low MHz frequencies and FFT-based spectral measurement.

\begin{figure}[b]
\center \label{figure1}
\includegraphics[width=3.25in]{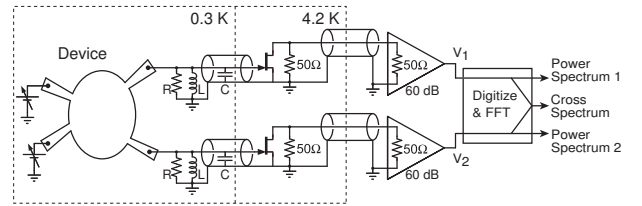}
\caption{\footnotesize{Block diagram of the two-channel noise
detection system, configured to measure the power
spectral densities and cross spectral density of current fluctuations
in a multi-terminal electronic device.}}
\end{figure}

Several factors make low-MHz frequencies a practical range for
low-temperature current noise measurement. This frequency range is
high compared to the $1/f$ noise corner in typical mesoscopic
devices. Yet, it is low enough that FFT-based spectral measurement
can be performed efficiently with a personal computer (PC) equipped
with a commercial digitizer. Key features of this FFT-based spectral
measurement are near real-time operation and sufficient frequency
resolution to detect spectral features of interest. Specifically,
the fine frequency resolution provides information about the
measurement circuit and amplifier noise at MHz, and enables
extraneous interference pick-up to be identified and eliminated.
These two features constitute a significant advantage over both
wide-band analog detection of total noise power, which sacrifices
resolution for speed, and swept-sine measurement, which sacrifices
speed for resolution.

The paper is divided as follows. A block diagram of the system is presented
in Section I. The amplification circuit is discussed in Section II.
Section III describes the data analysis procedure, including digitization and FFT processing.
A demonstration measurement of current noise in a quantum point contact (QPC)
is presented in Section IV. System performance is discussed in Section V.

\section{Overview of the system}

Figure~1 shows a block diagram of the two-channel noise detection
system, which is integrated with a  commercial $\Hethree$ cryostat
(Oxford Intruments Heliox $2^{\mathrm{VL}}$).  The system takes two
input currents and amplifies their fluctuations in several stages.
First, a parallel resistor-inductor-capacitor (RLC) circuit performs
current-to-voltage conversion at frequencies close to its resonance
at $f_o=(2\pi\sqrt{LC})^{-1}\approx2~\mathrm{MHz}$. Through its
transconductance, a high electron mobility transistor (HEMT)
operating at 4.2~K converts these voltage fluctuations into current
fluctuations in a $50~\Omega$ coaxial line extending from 4.2~K to
room temperature. A $50~\Omega$ amplifier with $60~\mathrm{dB}$ of
gain completes the amplification chain. The resulting signals $V_1$
and $V_2$ are simultaneously sampled at $10~\mathrm{MS}/\mathrm{s}$
by a two-channel digitizer (National Instruments PCI-5122) in a
3.4~GHz PC (Dell Optiplex GX280). The computer takes the FFT of each
signal and computes the power spectral density of each channel and
the cross spectral density.

\section{Amplifier}
\subsection{Design objectives}
A number of objectives have guided the design of the
amplification lines. These include:

\begin{enumerate}
\item Low amplifier input-referred voltage noise and current
noise.

\item Simultaneous measurement of both noise
at MHz and transport near dc.

\item Low thermal load.

\item Small size, allowing two amplification lines within the 52 mm bore
cryostat.
\item Maximum use of commercial components.

\item Compatibility with high magnetic fields.

\end{enumerate}

\subsection{Overview of Circuit}
Each amplification line consists of four circuit boards
interconnected by coaxial cable, as shown in the circuit schematic
in Fig.~2(a).  Three of the boards are located inside the $\Hethree$
cryostat. The resonant circuit board (labeled RES in Fig.~2(a)) is
mounted on the sample holder at the end of the 30~cm long coldfinger
that extends from the $\Hethree$ pot to the center of the
superconducting solenoid. The heat-sink board (SINK) anchored to the
$\Hethree$ pot  is a meandering line that thermalizes the inner
conductor of the coaxial cable. The CRYOAMP board at the
$4.2~\mathrm{K}$ plate contains the only active element operating
cryogenically, an Agilent ATF-34143 HEMT. The four-way SPLITTER
board operating at room temperature separates low and high frequency
signals and biases the HEMT. Each line amplifies in two frequency ranges, a 
low-frequency range below $\sim3~\mathrm{kHz}$ and a high-frequency range around 2
MHz.

\begin{figure}[t]
\center \label{figure2}
\includegraphics[width=3.25in]{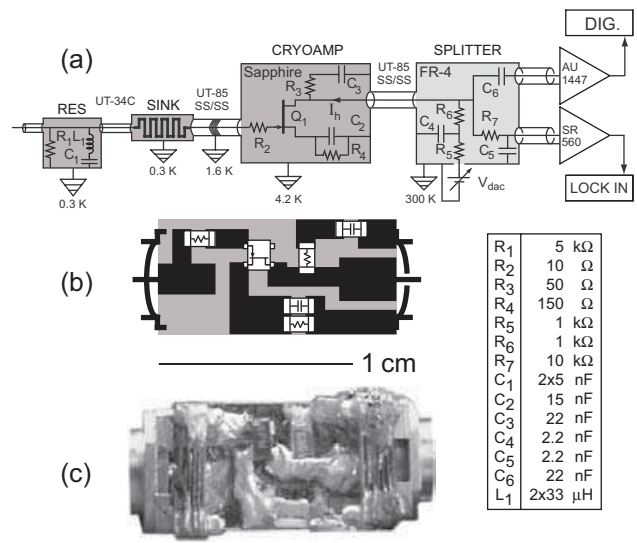}
\caption{\footnotesize {(a) Schematic diagram of each amplification
line. Values of all passive components are listed in the
accompanying table. Transistor $\mathrm{Q}_1$ is an Agilent
ATF-34143 HEMT. (b) Layout of the CRYOAMP circuit board.
Metal (black regions) is patterned by etching of thermally evaporated
Cr/Au on sapphire substrate. (c) Photograph of a CRYOAMP board. The
scale bar applies to both (b) and (c).}}
\end{figure}

 The low-frequency equivalent circuit
 is shown in Fig.~3(a): A  resistor ($R_1 = 5~\mathrm{k}\Omega$) to ground,
shunted by a capacitor ($C_1 = 10~\mathrm{nF}$), converts an input current
$i$ to a voltage on the HEMT gate. The HEMT amplifies this gate voltage by
$\sim -5~\mathrm{V}/\mathrm{V}$ on its drain, which connects to a
room temperature voltage amplifier at the low frequency port of the
SPLITTER board. The low-frequency voltage amplifier (Stanford Research
Systems model SR560) is operated in single-ended mode with ac coupling,
$100~\mathrm{V}/\mathrm{V}$ gain and bandpass filtering
($30~\mathrm{Hz}$ to $10~\mathrm{kHz}$). The bandwidth in this
low-frequency regime is set by the input time constant.

\begin{figure}[b]
\center \label{figure3}
\includegraphics[width=3.25in]{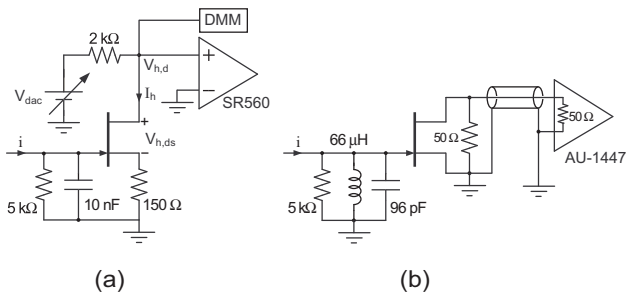}
\caption{\footnotesize {Equivalent effective circuits characterizing
the amplification line in the (a) low-frequency regime (up to
$\sim3~\mathrm{kHz}$), where it is used for differential conductance
measurements, and in the (b) high-frequency regime (few MHz), where
it is used for noise measurement.}}
\end{figure}

The high-frequency equivalent circuit is shown in Fig.~3(b).
The inductor $L_1 = 66~\mu$H dominates over $C_1$ and forms a parallel RLC tank
with $R_1$ and the capacitance $C \sim$ 96 pF of the coaxial line connecting to
the CRYOAMP board. Resistor $R_4$ is shunted by $C_2$ to enhance the
transconductance at the CRYOAMP board. The coaxial line extending from
$4.2~\mathrm{K}$ to room temperature is terminated on both sides by
$50~\Omega$. At room temperature, the signal passes through the
high-frequency port of the SPLITTER board to a $50~\Omega$ amplifier
(MITEQ AU-1447) with a gain of $60~\mathrm{dB}$ and a noise
temperature of $100~\mathrm{K}$ in the range
$0.01-200~\mathrm{MHz}$.

\subsection{Operating point}
The HEMT must be biased in saturation to provide voltage (transconductance) gain in 
the low (high) frequency range.  $R_4$, $R_5+R_6$ and supply
voltage $\vdac$ determine the HEMT operating point ($R_{1}$ grounds the HEMT gate at dc).
A notable difference in this design compared to
similar published ones regards the placement of $R_4$. In previous
implementations of similar circuits~\cite{Lee89, Lee93, Robinson04},
$R_4$ is a variable resistor placed outside the refrigerator and
connected to the source lead of $Q_1$ via a second coaxial line or
low-frequency wire. Here, $R_4$ is located on the CRYOAMP board to
simplify assembly and save space, at the expense of having full
control of the bias point in $Q_1$ ($R_4$ fixes the saturation value
of the HEMT current $\iq$). Using the I-V curves in
Ref.~28 for a cryogenically cooled ATF-34143, we
choose  $R_4=150~\Omega$ to give a saturation current of a few
mA. This value of saturation current reflects a compromise between
noise performance and power dissipation. As shown in Fig.~4, $Q_1$
is biased by varying the supply voltage $\vdac$ fed
at the SPLITTER board.  At the bias point indicated by a cross, the
total power dissipation in the HEMT board is  $\iq \vdsq+
\iq^2R_4=1.8~\mathrm{mW}$, and the input-referred voltage noise 
of the HEMT is $\sim0.4~\nVprH$.

\begin{figure}[t]
\center \label{figure4}
\includegraphics[width=2.75in]{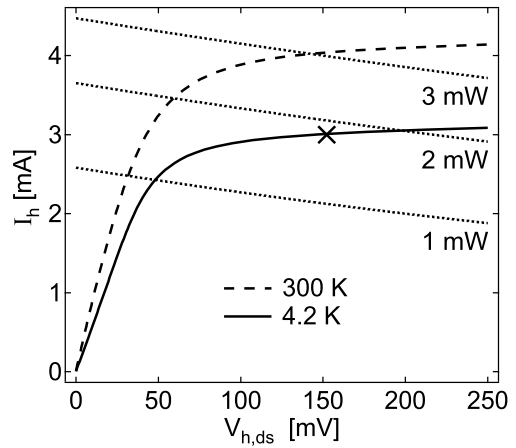}
\caption{\footnotesize{Drain current $\iq$ as a function of HEMT
drain-source voltage $\vdsq$, with the HEMT board at temperatures of
$300~\mathrm{K}$ (dashed) and $4.2~\mathrm{K}$ (solid). These curves
were obtained by sweeping the supply voltage $\vdac$ and measuring
drain voltage $\vdq$ with an HP34401A digital multimeter (see
Fig.~3(a)). From $\vdq$ and $\vdac$, $\iq$ and $\vdsq$ were then
extracted. Dotted curves are contours of constant power dissipation
in the HEMT board. The HEMT is biased in saturation (cross).}}
\end{figure}

\subsection{Passive Components}
Passive components were selected based on temperature stability,
size and magnetic field compatibility. All resistors (Vishay TNPW thin film)
are 0805-size surface mount. Their variation in
resistance between room temperature and $300~\mathrm{mK}$ is $<0.5\%$.
Inductor $L_1$ (two  $33~\mu\mathrm{H}$ Coilcraft 1812CS ceramic chip
inductors in series) does not have a magnetic core and is suited for operation at high
magnetic fields. The dc
resistance of $L_{1}$ is $26(0.3)~\Omega$ at 300(4.2)~K. With
the exception of $C_1$, all capacitors are 0805-size surface mount
(Murata COG GRM21). $C_1$ (two  $5~\mathrm{nF}$  American Technical Ceramics  700B
NPO capacitors in parallel)  is certified non-magnetic.

\subsection{Thermalization}
To achieve a low device electron temperature, circuit board
substrates must handle the heat load from the coaxial line. 
The CRYOAMP board must also handle the power
dissipated by the HEMT and $R_4$. Sapphire, having good thermal 
conductivity at low
temperatures~\cite{Pobell92} and excellent electrical insulation, is used for
the substrate in the RES, SINK and CRYOAMP boards. Polished blanks,
0.02" thick and  0.25" wide, were cut to lengths of 0.6" (RES and
CRYOAMP) or 0.8" (SINK) using a diamond saw. Both planar surfaces
were metallized with thermally evaporated Cr/Au (30/300~nm). Circuit
traces were then defined on one surface using a Pulsar toner
transfer mask and wet etching with  Au and Cr etchants (Transene
types TFA and 1020). Surface mount components were directly
soldered.

The RES board is thermally anchored to the sample holder with silver epoxy
 (Epoxy Technology 410E). The CRYOAMP (SINK) board is thermalized to
 the $4.2~\mathrm{K}$ plate  ($\Hethree$ pot) by a copper braid soldered
 to the back plane.

Semirigid stainless steel coaxial cable (Uniform Tube UT-85 SS-SS)
is used between the SINK and CRYOAMP boards, and between the CRYOAMP board
and room temperature. Between the RES and SINK boards, smaller coaxial cable
(Uniform Tube UT-34 C) is used to conserve space.

With this approach to thermalization, the base temperature of the
$^3\mathrm{He}$ refrigerator is 290~mK with a hold time of $\sim
45$~h. As demonstrated in Section IV, the electron base temperature
in the device is also 290 mK.

\section{Digitization and FFT Processing}

The amplifier outputs $V_1$ and $V_2$ (see Fig.~1) are sampled
simultaneously using a commercial digitizer (National Instruments PCI-5122)
with 14-bit resolution at a rate $f_{s}=10~\mathrm{MS/s}$.
To avoid aliasing\cite{Oppenheim89} from the broad-band amplifier background,
$V_1$ and $V_2$ are frequency limited to below the Nyquist frequency of
$5~\mathrm{MHz}$ using 5-pole Chebyshev low-pass filters, built
in-house from axial inductors and capacitors with values specified
by the design recipe in Ref.~31. The filters have a measured
half power frequency of $3.8~\mathrm{MHz}$, $39~\mathrm{dB}$
suppression at $8~\mathrm{MHz}$ and a pass-band ripple of
$0.03~\mathrm{dB}$.

While the digitizer continuously stores acquired data into its
memory buffer (16 MB per channel), a software program processes the data
from the buffer in blocks of $M=10,368$ points per channel. $M$ is
chosen to yield a resolution bandwidth $f_{s}/M\sim~1~\mathrm{kHz}$,
and to be factorizable into powers of two and three to maximize the efficiency of the FFT
algorithm.

Each block of data is processed as follows. First,  $V_1$ and $V_2$  are
multiplied by a Hanning window $W_{H}[m]=\sqrt{2/3}(1-\cos(2\pi m / M))$
to avoid end effects\cite{Oppenheim89}. Second, using the FFTW
package~\cite{FFTW98}, their FFTs are calculated:
\begin{equation*}
\widetilde{V}_{1(2)}[f_{n}]=
\sum_{m=0}^{M-1}W_{\mathrm{H}}[m]V_{1(2)}(t_{m})e^{-i2\pi f_{n}
t_{m}},
\end{equation*}
where $t_{m}=m/f_{s}$, $f_{n}=(n/M)f_{s}$, and $n={0,1,...,M/2}$.
Third, the power spectral densities $P_{1,2} =
2|\widetilde{V}_{1,2}|^2/(M f_{s})$ and the cross spectral density
$X=2(\widetilde{V}_{1}^{*}\cdot\widetilde{V}_{2})/(M
f_{s})=X_R+iX_I$ are computed.

As blocks are processed, running averages of $P_{1}$, $P_{2}$, and
$X$ are computed until the desired integration time
$\tau_\mathrm{int}$ is reached. With the 3.4~GHz computer and the
FFTW algorithm, these computations are carried out in nearly
real-time: it takes $10.8~\mathrm{s}$ to acquire and process
$10~\mathrm{s}$ of data.

\section{Measurement: QPC current noise}
In this section, the  system is demonstrated with measurements
of current noise in a quantum point contact (QPC). Specifically, the partition
noise $\Sipart$ is measured as a function of QPC source-drain bias $\vsd$:
\begin{equation}
\Sipart(\vsd)=S_I(\vsd)-4 k_B \te g(\vsd).
\end{equation}
Here, $S_I$ is the total QPC current noise spectral density
without extraneous noise ($1/f$, random telegraph, pick-up),
$k_B$ is the Boltzmann constant, $\te$ is the electron
temperature, $g(\vsd)=dI/d\vsd$ is the bias-dependent QPC
differential conductance, and $I$ is the current through the QPC.

\subsection{Device and Setup}
The QPC is defined by two electrostatic gates on the surface of a
$\mathrm{GaAs}/\mathrm{Al}_{0.3}\mathrm{Ga}_{0.7}\mathrm{As}$
heterostructure grown by molecular beam epitaxy. The two-dimensional
electron gas (2DEG) $190~\mathrm{nm}$ below the surface has a
density $1.7\times 10^{-11}~\mathrm{cm}^{-2}$ and mobility
$5.6\times10^6~ \mathrm{cm}^2/\mathrm{Vs}$. The QPC conductance is
controlled by negative voltages $V_{\mathrm{g1}}$ and
$V_{\mathrm{g2}}$ applied to the electrostatic gates.

The QPC is connected to the system as shown in the inset of Fig.~5.
The two amplification lines are connected to the same reservoir of
the QPC. In this case, the two input RLC tanks effectively become a single tank
with resistance $R'\approx2.5~\mathrm{k}\Omega$, inductance
$L'\approx33~\mu\mathrm{H}$ and capacitance
$C'\approx192~\mathrm{pF}$. The QPC current noise couples to both 
amplification lines and thus can be extracted from either
the single channel power spectral densities or the cross spectral density.
The latter has the technical advantage of rejecting any noise not common
to both amplification lines. It is used to extract $\Sipart$ in the remainder
of this section.

A $25~\mu\mathrm{V}_{\mathrm{rms}}$, $430~\mathrm{Hz}$ excitation $\vac$
is applied to the other QPC reservoir and used for
lock-in measurement of $g$. A dc bias voltage $\vdc$ is also applied to
generate a finite $\vsd$. $\vsd$ deviates from $\vdc$
due to the resistance in-line with the QPC,
which is equal to the sum of $R_{1}/2$ and ohmic contact resistance $R_{s}$.
$\vsd$ could in principle be measured by the traditional four-wire technique. This  
would require additional low-frequency wiring,  as well as filtering to prevent extraneous pick-up
and room-temperature amplifier noise from coupling to the noise measurement circuit.
For technical simplicity, here $\vsd$ is obtained by numerical integration of the measured bias-dependent $g$:
\begin{equation}
\vsd=\int_0^{\vdc}\frac{dV}{1+(R_{1}/2+R_s)g(V)}
\end{equation}

\subsection{Measurement}

\begin{figure}[t]
\center \label{figure5}
\includegraphics[width=2.75in]{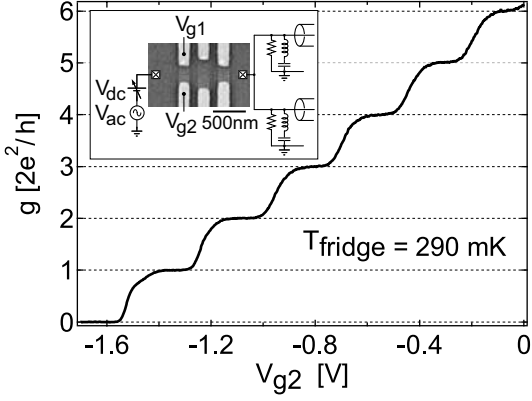}
\caption{\footnotesize{Inset: Setup for detection of QPC current
noise and electron micrograph of a device identical in design to the one used.
The QPC is defined by negative voltages $\vgone$ and
$\vgtwo$ applied on two facing gates. All other gates in the
device are grounded. Main: linear conductance $g(\vsd=0)$ as
a function of $\vgtwo$ at 290~mK, measured using
amplification line 1. $\vgone=-3.2~\mathrm{V}$.}}
\end{figure}

Figure~5 shows linear conductance $g(\vsd=0)$ as a function of $\vgtwo$, at a
fridge temperature $\Tfridge=290~\mathrm{mK}$ (base temperature). Here,
$g$ was extracted from lock-in measurements using amplification line 1.
As neither the low frequency gain of amplifier 1 nor $R_s$ were
known precisely beforehand, these parameters were calibrated by
aligning the observed conductance plateaus to the expected multiples
of $\tgq$. This method yielded a low frequency gain $-4.6~\mathrm{V}/\mathrm{V}$
and $R_s=430~\Omega$.

\begin{figure}[b]
\center \label{figure6}
\includegraphics[width=2.75in]{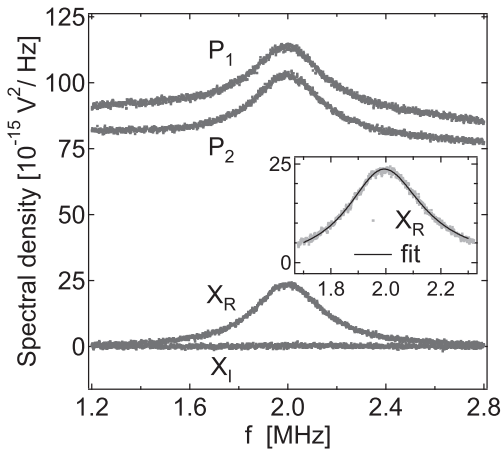}
\caption{\footnotesize{Power spectral densities $P_1$ and $P_2$, and
real and imaginary parts $X_R$ and $X_I$ of the cross spectral
density, at base temperature and with the QPC pinched off $(g=0)$,
obtained from noise data acquired for $\tau_\mathrm{int}=20~\mathrm{s}$. Inset:
expanded view of $X_R$ near resonance, along with a fit using
Eq.~(3) over the range $1.7$ to $2.3~\mathrm{MHz}$.}}
\end{figure}

Figure~6 shows $P_1$, $P_2$, $X_R$, and $X_I$ as a function of
frequency $f$,
at base temperature and with the QPC pinched off ($g=0$). $P_{1(2)}$
shows a peak at the resonant frequency of the RLC tank, on top of a
background of approximately
$85(78)\times10^{-15}~\mathrm{V}^2/\mathrm{Hz}$. The background in
$P_{1(2)}$ is due to the voltage noise $S_{V,1(2)}$ of amplification
line 1(2) ($\sim 0.4~\nVprH$). The peak results from thermal noise
of the resonator resistance and current noise ($S_{I,1}+S_{I,2}$)
from the amplifiers.  $X_R$ picks out this peak and
rejects the amplifier voltage noise backgrounds. The inset zooms in
on $X_R$ near the resonant frequency. The solid curve is a best-fit to the
form

\begin{equation}
X_R(f)=\frac{X_{R}^{0}}{1+(f^2-f_o^2)^2/(f \Delta
f_{3\mathrm{dB}})^2},
\end{equation}
corresponding to the lineshape of white noise band-pass filtered by
the RLC tank. The fit parameters are the peak height $X_R^0$, the
half-power bandwidth $\Delta f_{3\mathrm{dB}}$ and the peak
frequency $f_o$. Power spectral densities $P_{1(2)}$ can
be fit to a similar form including a fitted background term:

\begin{equation}
P_{1(2)}(f)=P_{1(2)}^{\mathrm{B}}+
\frac{P_{1(2)}^{0}}{1+(f^2-f_o^2)^2/(f \Delta f_{3\mathrm{dB}})^2},
\end{equation}

\subsection{Noise measurement calibration}
In order to extract $\Sipart$ from $X_{R}(f)$, the noise measurement system
must be calibrated \textit{in situ}.  An effective circuit with noise sources
is defined for this purpose and shown in Fig.~7.  Within this circuit model, 
$\Sipart$ is given by: \begin{equation}
\Sipart = \left(\frac{X_R^0}{\gx^2} - 4 k_B
\te\Reff\right)\left(\frac{1+g R_s}{\Reff}\right)^2
\end{equation}
Here,  $\gx=\sqrt{G_1 G_2}$ is the cross-correlation gain and
$\Reff=2 \pi f_o^2 L' / \Delta f_{\mathrm{3dB}}$ is the total effective 
resistance parallel to the tank~\cite{Detail01}.

Calibration requires assigning values for $R_{s}$,  $\te$, and $\gx$.
While the value $R_{s}=430~\Omega$ is known from the conductance measurement,
$\gx$ and $\te$ are calibrated from thermal noise measurements. The
procedure demonstrated in Fig.~8 stems from the relation 
$X_R^0=4 k_B \te \Reff \gx^2$~\cite{Detail02} valid at $\vsd=0$.  

First, $X_{R}(f)$ is measured over $\tint=30~\mathrm{s}$ for various $\vgtwo$ settings at
each of three elevated fridge temperatures ($\Tfridge=3.1,~4.2,~\mathrm{and}~
5.3~\mathrm{K}$). $X_R^0$ and $\Reff$ are extracted from fits to $X_{R}(f)$ using 
Eq.~(3)  and plotted parametrically (open markers in Fig.~8(a)). A linear fit (constrained to
pass through the origin) 
to each parametric plot gives the slope $dX_R^0/d\Reff$ at each temperature, 
equal to $4 k_B \te \gx^2$. Assuming $\te = \Tfridge$ at these temperatures,
$\gx=790~\mathrm{V}/\mathrm{V}$ is  extracted from a linear fit
to $dX_R^0/d\Reff(\Tfridge)$, shown in Fig.~8(b).

\begin{figure}[b]
\center \label{figure7}
\includegraphics[width=3.25in]{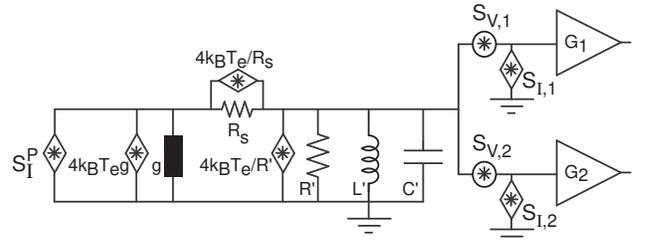}
\caption{\footnotesize{Circuit model used for extraction of the 
QPC partition noise $\Sipart$. $G_{1(2)}$ is the 
voltage gain of amplification line 1(2) between HEMT gate and digitizer input.
}}
\end{figure}

\begin{figure}[t]
\center \label{figure8}
\includegraphics[width=2.75in]{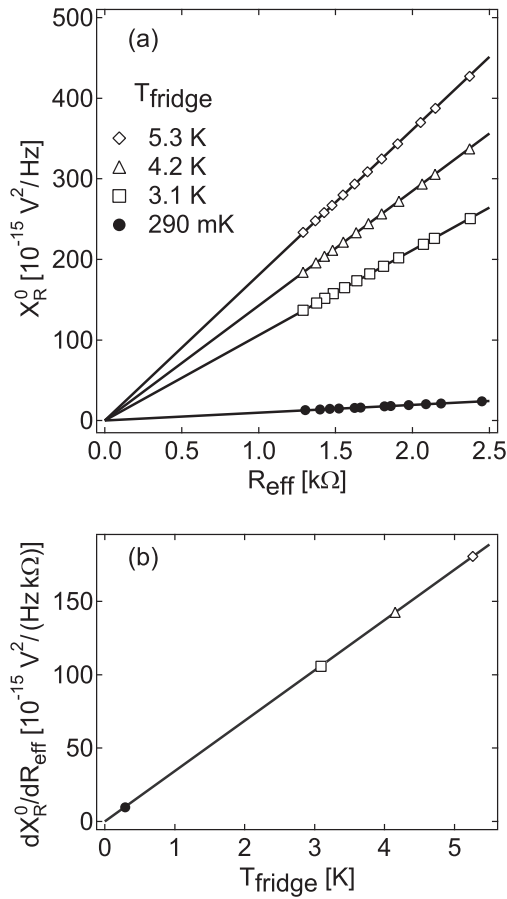}
\caption{\footnotesize{
Calibration by noise thermometry 
of the electron temperature $\te$ at base fridge temperature  and  the
cross-correlation gain $\gx$. (a) $X_R^0$ as function of $\Reff$
(both from fits to $X_{R}(f)$ using Eq.~(3)), at base
(solid circles) and at three elevated fridge temperatures (open
markers). Solid lines are linear fits constrained to the origin. (b)
Slope $dX_R^0/d\Reff$ (from fits in (a)) as a function of $\Tfridge$.
Solid line is a linear fit (constrained to the origin) of  $dX_R^0/d\Reff$
at the three elevated temperatures (open markers).}}
\end{figure}

Next, the base electron temperature is calibrated from 
a parametric plot of  $X_R^0$ as a function of $\Reff$ obtained from similar 
measurements at base temperature (solid circles in Fig.~8(a)). From the fitted slope
$dX_R^0/d\Reff$  (black marker in Fig.~8(b)) and using
the calibrated $\gx$, a value $\te=290~\mathrm{mK}$ is obtained. This
suggests that electrons are well thermalized to the fridge.

\subsection{QPC partition noise}

Following the calibration, $\Sipart(\vsd)$ is extracted as follows.
$X_{R}(f)$  and  $g$ are simultaneously measured ($\tint=60~\mathrm{s}$)
at fixed $\vgtwo$ as a function of $\vsd$ between $-150~\mu\mathrm{V}$ 
and $+150~\mu\mathrm{V}$. At each $\vsd$ setting,  $X_{R}^{0}$ and $\Reff$ 
are obtained from fits of $X_{R}(f)$ to Eq.~(3), and used with the measured $g$ to 
extract $\Sipart(\vsd)$ from Eq.~(5). 

Demonstration measurements of
$\Sipart(\vsd)$ are shown in Fig.~9. Open markers superimposed on
the linear conductance trace in Fig.~9(a) indicate $\vgtwo$ settings
giving $g(\vsd=0) \approx 0$, 0.5, 1, 1.5, and $2\ttgq$.
The corresponding noise data are shown in Fig.~9(b). At 0, 1 and
$2\ttgq$, where the QPC is either pinched off or on a linear conductance
plateau, $\Sipart$ shows little dependence on bias, in contrast with
the $|\vsd|$ dependence observed when $g \approx$ 0.5 and $1.5\ttgq$.
This behavior is consistent with earlier experiments~\cite{Reznikov95,Kumar96} and
theory~\cite{Lesovik89,Buttiker90} of shot noise in a QPC.

\begin{figure}[t]
\label{figure9}
\includegraphics[width=3.25in]{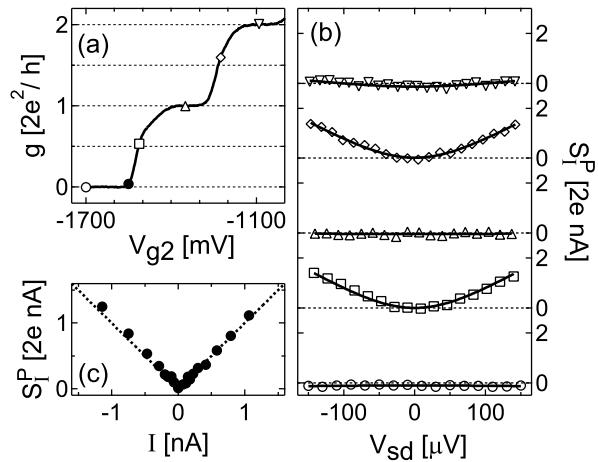}
\caption{\footnotesize{(a) Linear conductance $g(\vsd=0)$ as a function of $\vgtwo$.
Open and solid markers indicate $\vgtwo$ settings for the noise
measurements shown in (b) and (c), respectively. (b) QPC partition noise
$\Sipart$ as a function of $\vsd$, for
conductances near 0, 0.5, 1, 1.5, and 2~$\ttgq$. Solid lines are
fits to Eq.~(6) using $\NF$ as the only fitting parameter. (c)
$\Sipart$ as a function of dc current $I$ with the
QPC near pinch-off. The dotted line represents full shot noise
$\Sipart=2e|I|$.}}
\end{figure}

Within mesoscopic scattering theory~\cite{Blanter00,Martin05}, where transport is
described by transmission coefficients $\tauns$ ($n$ is the sub-band
index and $\mathrm{\sigma}$ denotes spin), $\Sipart$ is given by
\begin{equation}
\Sipart(\vsd) = 2\frac{2e^2}{h}\NF\left[e\vsd\coth\left(\frac{e\vsd}{2\kb\te}\right)-2\kb\te\right],
\end{equation}
with a \textit{noise factor} $\NF =
\frac{1}{2}\sum\tauns(1-\tauns)$. This equation is strictly valid
for constant transmission coefficients across the bias window.
At low-temperatures and for the spin-degenerate case, 
$\NF$ is zero at multiples of $\tgq$ and reaches a maximum value of 0.25
at odd multiples of $0.5\ttgq$. Fits to the $\Sipart(\vsd)$ data in
Fig.~9(b) using the form of Eq.~(6) are shown as solid curves, 
with $\te=290~\mathrm{mK}$ and
best-fit $\NF$ values of 0.00, 0.20, 0.00, 0.19, and 0.03 for
$g\approx$ 0, 0.5, 1, 1.5, and $2\ttgq$, respectively. The
deviation of the best-fit $\NF$ from 0.25 near 0.5 and 1.5$\ttgq$ is
discussed in detail in Ref.~37.

A measurement of $\Sipart$ as a function of $I$ with
the QPC barely open (solid marker in Fig.~9(a)) is shown 
in Fig.~9(c).  In this regime, full shot noise 
 $\Sipart=2e|I|$ is observed. This is consistent with scattering theory and
with recent measurements on mesoscopic tunnel barriers
free of impurities, localized states and 1/f noise\cite{Chen06}. 

\section{System Performance}

\begin{figure}[t]
\center \label{figure10}
\includegraphics[width=2.75in]{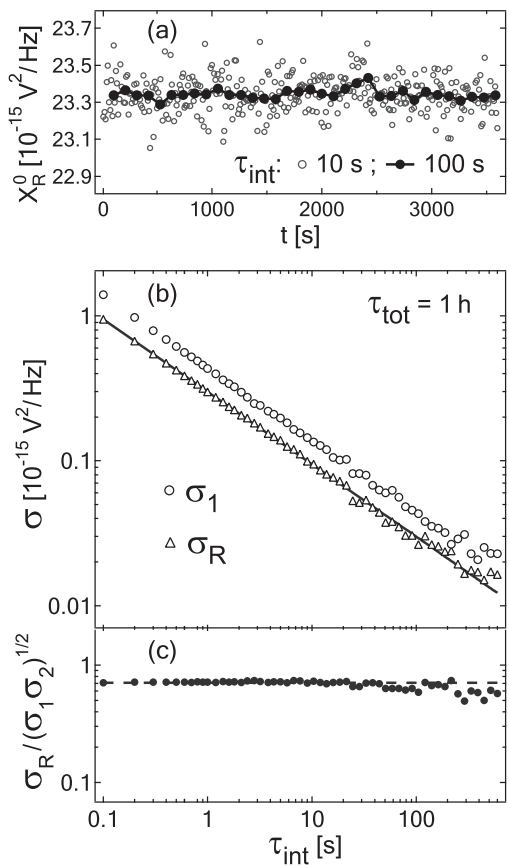}
\caption{\footnotesize {(a) $X_R^0$ as a function of time $t$, for
$\tau_{\mathrm{int}}$ of $10~\mathrm{s}$ (open circles) and
$100~\mathrm{s}$ (solid circles). (b) Standard deviations $\sigma_1$ and
$\sigma_R$  as a function of
$\tau_{\mathrm{int}}$. Solid line is a fit to $\sigma_R$ of the form
$C \cdot \tau_{\mathrm{int}}^{-1/2}$, with best-fit value $C=0.30\times
10^{-15}~\mathrm{s}^{1/2}\mathrm{V}^2/\mathrm{Hz}$. (c)
$\sigma_R/\sqrt{\sigma_1 \sigma_2}$ as a function of
$\tau_{\mathrm{int}}$. The dashed line is a constant $1/\sqrt{2}$.
}}
\end{figure}

The resolution in the estimation of current noise spectral density from
one-channel and two-channel measurements is determined experimentally
in this final section. Noise data are first sampled over a total
time  $\tau_{\mathrm{tot}}=1~\mathrm{h}$, with the QPC at base temperature and pinched off.
Dividing the data in segments of time length $\tau_{\mathrm{int}}$, calculating the
power and cross spectral densities for each segment, and fitting
with Eqs. (3) and (4) gives a sequence of
$\tau_{\mathrm{tot}}/\tau_{\mathrm{int}}$ peak heights for each of
$P_1$, $P_2$ and $X_R$. Shown in open (solid) circles in Fig.~10(a) is
$X_R^0$ as a function of time $t$ for $\tau_{\mathrm{int}}=10(100)~\mathrm{s}$.
The  standard deviation $\sigma_{R}$ of  $X_R^0$ is $1(0.3)\times10^{-16}~\mathrm{V}^2/\mathrm{Hz}$.
The resolution  $\delta S_I$  in current noise spectral density is given
by $\sigma_{R}/(\gx^{2} \Reff^2)$  (see Eq.~(5)).
For  $\tau_{\mathrm{int}}=10~\mathrm{s}$,  $\delta S_I=2.8\times10^{-29}~\mathrm{A}^2/\mathrm{Hz}$,
which  corresponds to full shot noise $2eI$ of $I\sim 100~\mathrm{pA}$.

The effect of integration time on the resolution is determined
by repeating the analysis for different values of $\tau_{\mathrm{int}}$.
Fig.~10(b) shows the standard deviation $\sigma_1$ ($\sigma_R$) of
$P_1^0$ ($X_R^0$) as a function of $\tau_{\mathrm{int}}$. The
standard deviation $\sigma_2$ of $P_2^0$, not shown, overlaps
closely with $\sigma_1$. All three standard deviations scale as
$1/\sqrt{\tau_{\mathrm{int}}}$, consistent with the Dicke radiometer
formula~\cite{Dicke46} which applies when measurement
error results only from finite integration time, i.e., it is purely
statistical. This suggests that, even for the longest segment length of $\tint=10~\mathrm{min}$,
the measurement error is dominated by statistical error and not by instrumentation drift on the scale of 1~h.

Figure~10(c) shows $\sigma_{\mathrm{R}}/\sqrt{\sigma_{1}\sigma_{2}}$ as a function of
$\tau_{\mathrm{int}}$. This ratio gives the fraction by which, in the present measurement
configuration, the statistical error in current noise spectral density estimation from $X_{R}^{0}$ is lower than the error 
in the estimation from either $P_{1}^{0}$ or $P_{2}^{0}$ alone.  The geometric mean in the denominator accounts for 
any small mismatch in the gains $G_{1}$ and $G_{2}$.  In theory, and in the absence of drift, this ratio is independent of
$\tint$ and equal to $1/\sqrt{2}$ when the uncorrelated amplifier voltage noise $(S_{V,1(2)})$ dominates over the 
noise common to both amplification lines. The ratio would be unity when the correlated noise
dominates over $S_{V,1(2)}$. 

The experimental $\sigma_{\mathrm{R}}/\sqrt{\sigma_{1}\sigma_{2}}$ is close to 
$1/\sqrt{2}$ (dashed line).  This is consistent with  the spectral density data in Fig.~6, which shows
that the backgrounds in $P_{1}$ and $P_{2}$ are approximately three times larger than the 
cross-correlation peak height.  The ratio deviates slightly below $1/\sqrt{2}$ at the largest $\tint$ values.
This may result from enhanced sensitivity to error in the substraction of the
$P_{1(2)}$ background at the longest integration times. 
 
A similar improvement relative to estimation from either $P_{1}^{0}$ or  $P_{2}^{0}$ alone would
also result from estimation with a weighted average $(P_1^0/G_1^2 + P_2^0/G_2^2)G_X^2/2$. 
The higher resolution attainable from two channel measurement relative to
single-channel measurement in this regime has been previously exploited in noise measurements in the 
kHz range~\cite{Kumar96,Glattli97,Sampietro99}.

\section{Conclusion}
We have presented a two-channel noise detection system measuring auto- and
cross-correlation of current fluctuations near $2~\mathrm{MHz}$ in electronic
devices at low temperatures.  The system  has been implemented in a $^3\mathrm{He}$
refrigerator where the base device electron temperature, measured by noise thermometry,
is $290~\mathrm{mK}$. Similar integration with a $\Hethree$ -$\Hefour$
dilution refrigerator  would enable noise measurement at temperatures of  tens of mK.

We thank  N.~J.~Craig, J.~B.~Miller, E.~Onitskansky, and
S.~K.~Slater for device fabrication. We also thank H.-A.~Engel, D.~C.~Glattli,
P.~Horowitz, W.~D.~Oliver, D.~J.~Reilly, P.~Roche, A.~Yacoby, Y.~Yamamoto for valuable
discussion, and B.~D'Urso, F.~Molea and H.~Steinberg for technical assistance.
We acknowledge support from NSF-NSEC, ARDA/ARO, and Harvard University.


\begin{thebibliography}{39}
\bibitem[{\citenamefont{Detail02}(05)}]{Detail02}
The full expression within the 
circuit model is $X_R^0 =
(4 k_B \te \Reff + (S_{I,1}+S_{I,2})\Reff^2) \gx^2$. The linear
dependence of $X_R^0$ on $\Reff$ observed in Fig.~8(a) demonstrates
that the quadratic term from amplifier current noise is negligible.
\end{thebibliography}
\end{document}